\documentclass[moor,nonblindrev]{informs1} 

\usepackage{natbib}
 \NatBibNumeric
 \def\bibfont{\small}%
 \bibpunct[, ]{[}{]}{,}{n}{}{,}%
\usepackage{amssymb}
\usepackage[colorlinks=true,breaklinks=true,bookmarks=true,urlcolor=blue,
     citecolor=blue,linkcolor=blue,bookmarksopen=false,draft=false]{hyperref}
 
\usepackage{nomencl, enumerate}
\makenomenclature
\usepackage{titlesec} 
\titleformat{\subsection}[runin]{\normalfont\bfseries}{\thesubsection.}{0pt}{}

\newcommand{\real}{\mathbb{R}}
\newcommand{\expt}{\mathbf{E}}
 
\def\EMAIL#1{\href{mailto:#1}{#1}}
\def\URL#1{\href{#1}{#1}}         

\TheoremsNumberedThrough     

\EquationsNumberedThrough    

\renewcommand{\theARTICLETOP}{}
\begin{document}
\RUNAUTHOR{Luo and Saigal}

\RUNTITLE{A Note on the Multi-Agent Contracts in Continuous Time}

\TITLE{A Note on the Multi-Agent Contracts in Continuous Time}
\ARTICLEAUTHORS{%
\AUTHOR{Qi Luo}
\AFF{Department of Industrial and Operations Engineering, University of Michigan. \EMAIL{luoqi@umich.edu}, \URL{}}
\AUTHOR{Romesh Saigal}
\AFF{Department of Industrial and Operations Engineering, University of Michigan. \EMAIL{rsaigal@umich.edu}, \URL{}}
} 

\ABSTRACT{%
Dynamic contracts with multiple agents is a classical decentralized decision-making problem with asymmetric information. In this paper, we extend the single-agent dynamic incentive contract model in continuous-time to a multi-agent scheme in finite horizon and allow the terminal reward to be dependent on the history of actions and incentives. We first derive a set of sufficient conditions for the existence of optimal contracts in the most general setting and conditions under which they form a Nash equilibrium. Then we show that the principal's problem can be converted to solving Hamilton-Jacobi-Bellman (HJB) equation requiring a static Nash equilibrium. Finally, we provide a framework to solve this problem by solving partial differential equations (PDE) derived from backward stochastic differential equations (BSDE).   
}%

\KEYWORDS{Principal Multi-Agent Problem, Continuous-Time Approach, Kakutani Fixed Point Theorem, Nash Equilibrium, Hamilton-Jacobi-Bellman Equation}
\HISTORY{\noindent Current version is uploaded on September 16, 2017.}

\maketitle
%
%
%
%
%
%
\section{Introduction} \label{S1:intro} 
\noindent Dynamic contract theory has been used for many practical problems ranging from corporate finance to strategic behavior in politics to organizational design. A principal (``She'') is a person or entity who constructs a contract and an agent (``He'') decides to possibly accept it. A dynamic contact defines the payoff received by the agent depending on his exerted effort during each period the contract is in effect. The principal needs to characterize all admissible contracts and implement one that maximizes her expected revenue from the cash flow. Some early work on principal-agent problem in repeated games includes the two-period problem  of Rogerson (1985), infinite repeated moral hazard problem of Spear and Srivastava (1987), renegotiation in long-term contract problem of Hart and Tirole (1988), and asymptotic efficiency contract of Fudenberg, Holmstrom, Milgrom (1990) (\cite{rogerson1985repeated,spear1987repeated,hart1988contract,fudenberg1990short}). In this paper, we focus on the dynamic incentive contract (i.e. moral hazard) problem where agents actions are not observable by the principal. In this case, the first-best contract is not admissible since a contract conditioned on agent's verifiable actions cannot be implemented. This could be because it may be too costly, impossible or difficult to monitor the agent's actions directly.  In such a situation, the principal can only offer an incentive contract which rewards the agent according to the level of revenue he generates and she receives. 
 
In the groundbreaking work, Sannikov (2008) developed a continuous-time approach to the dynamic principal-agent problem, using some ideas of Holmstrom and Milgrom who described the continuous-time output process as a Brownian motion about twenty years ago(\cite{sannikov2008continuous,holmstrom1987aggregation}). As is the case in a discrete model, an agent only cares about his continuation value resulting from his current action. The principal receives the payoff determined by an output process whose drift term is determined by the agent's action and the diffusion  term is assumed to be a Brownian Motion. The value functions of both the principal and the agent, in this continuous setting, are Martingales with respect to the filtration of the Brownian motion. Both the agent and the principal solve two embedded dynamic programming problems to determine the target continuation values. This continuous-time framework models the dynamics of the process analytically and bypasses the problems that arise due to the curse of dimensionality which results from the enormous state space of a history-dependent contract. This is done by choosing the agents continuation value as the state variable for the principals problem, thus avoiding the explosion of the state space \cite{sannikov2008continuous}. This continuous approach can also model different and correlated stochasticity which is difficult, if not impossible, to model using the discrete approach. For instance, the principal can control the initial condition, drift terms, diffusion terms, and solve the contract in time-variant interest rate setting \cite{bergemann2015dynamic,cvitanic2015dynamic,cvitanic2016moral,williams2015solvable}. 

Though both discrete-time and continuous-time methods find the optimal policy by solving the Bellman and the Hamilton-Jacobi-Bellman (HJB) equations respectively, each has strengths and weaknesses with regard to the computational efficiency. Discrete-time model can be reformulated with objective function and constraints linear in probability, and thus can be converted to linear programming (LP) subproblems. Continuous-time method solves BSDE so it is harder to program computers to solve (we show in this note that the multi-agent problem  can be solved by a series of parabolic nonlinear PDE). However, the dimensionality of each linear program grows exponentially with the time horizon and this curse of dimensionality vanishes in the continuous-time models. 

Discrete-time dynamic multi-agent contract problems have been relatively well studied, but only recently researchers have focused the optimal strategy for the continuous-time problem in a multi-agent setting. A linear extension of the drift term in this multi-agent setting has been looked at by Thakur with the aim of finding the optimal size of team (number of agents) when dividing work between them \cite{thakur2015continuous}. In the trivial linear setting, the only direct interaction between agents is through the cost function which is assumed to be a decreasing function of the team size. To the best of our knowledge, the pioneering paper that discusses the Nash Equilibrium as a concept to connect the decisions of competitive interacting agents in this general setting is by Elie and Possamai \cite{elie2016tale,elie2016contracting}. The characterization of the Nash equilibrium to the multidimensional BSDE as well as the existence of such equilibrium in this general setting are both discussed with the assumption that the mapping from agents' actions to drift term of the principal's output is uniformly Lipschitz continuous. However, the authors presume the existence of equilibrium, and disregard the complex issue of simultaneously finding the equilibrium and optimizing admissible contracts in the multi-agent setting.    

Our contribution to the contract theory literature include proving a set of sufficient conditions for the existence of optimal contracts in continuous-time multi-agent problem, as well as providing the algorithm for solving the optimal contracts in the most general setting. Compared to previous work, we generalize the forms of output processes, utility and aggregate functions, expanding the application domains of this methodology. 

The rest of this paper is organized as follows. We first describe the setting of a continuous-time single-principal multi-agent problem in Section \ref{S2:setting}. In accordance with procedures to solve optimal dynamic contract, we first characterize the agents' problem as a static subproblem in Section \ref{S3:SS1agent}. Multi-agent requires more effort to prove the existence of sub-game perfect Nash equilibrium in Section \ref{S4:equilibrium}.  Next we formulate the principal's problem as a HJB equation based on specified agents' actions at equilibrium in Section \ref{S3:SS3principal} and in Section \ref{S3:solve} we give some insight into solving the optimal contract. Finally the Section \ref{S5:conclusion} presents our conclusions and some possible future directions.   

\section{Setting}\label{S2:setting}
\noindent There is one principal (subscripted as ``$P$'') and $n$ agents, signing a contract simultaneously at time $t = 0$. 
Each agent is indexed by $i$, $i \in \{1, 2, \dots n \} $ and (subscripted as ``$i$''). A contract signed between the principal and an agent $i$ specifies the compensations $c_i(t)$ this agent will receive by putting effort $a_i(t)$ in working for the principal at time $t \in [0, T]$. The vectors of $n$ agents' actions and compensations are denoted as $\pmb{a}(t)$ and $\pmb{c}(t)$ respectively. There is a terminal payoff $\Phi_i$ at the end of time horizon $T < \infty$. We argue that the infinite horizon is a special case where $T=\infty$ and there is no terminal payoff. In this case, an analogous approach from this paper still works but with additional transversality conditions to guarantee the existence of optimal contracts on the interval including $\infty$. On the other hand, the state space at each $t$ is reduced because of removing boundary conditions. In continuous-time model, the order of realizing payments and determining the effort levels becomes irrelevant. 

We allow the terminal payoff for $n$ agents $\pmb{\Phi}$ to be history-dependent, i.e. $\pmb{\Phi}$ is a vector of functions of $\{\pmb{a}(t), \pmb{c}(t)\}_{0 \leq t \leq T}$. For example, the terminal payoff for agent $i$ denoted as $\Phi_i$ can be a possibly a nonlinear function of $Z_i(T)$ where $\pmb{Z}(t)$ is the sum total of the agents' actions from $0$ to $t$ 
\begin{align}
	\pmb{Z}(t) &= \int_{0}^{t} \pmb{a}(s) d s. 
\end{align}
or of some other functions of the actions and/or incentives. In the general case we represent the dynamics of $\pmb{Z}$ for $i=1,\cdots,n$ as
\begin{align}
dZ_i(t)=r\mu_{Z_i}(\pmb{a}(t),\pmb{c}(t))dt +r\pmb{\sigma}_{Z_i}d\pmb{B}_Z(t)
\label{eqn11}
\end{align}
where $\mu_{Z_i},\pmb{\sigma}_{Z_i}$ are deterministic functions of appropriate dimension, and $\pmb{B}_Z$ is an $m$ dimensional vector of independent Brownian motions. 

The expected utility the agent $i$ draws from a contact is $U_i$, and the total utility the principal obtains from signing contracts with $n$ agents is $U_P$. We assume that the instantaneous payoff received by the agent $i$ at time $t\in[0,T]$ is
\begin{align}
u_i(\pmb{a}(t),c_i(t))
\label{eq1}
\end{align}

\noindent  where the principal's decision variable is the compensation $c_i(t)$ in domain $\mathcal{C}_i \subseteq \real$, and the agent $i$'s decision variable is his effort level $a_i(t)$ in domain $\mathcal{A}_i \subseteq \real$. Both domains are agent-dependent, and their Cartesian products are denoted as $\mathcal{C}$ and $\mathcal{A}$. In an uncertain environment, the output processes associated with $n$ agents' actions is a vector $\pmb{X}(t) = [X_1(t), \dots, X_n(t)]$. Each $X_i(t)$ follows the process: 
\begin{align}
	d X_i(t) = f_i(\pmb{a}(t)) d t + \pmb{\sigma}_i d \pmb{B}(t)
    \label{eq2}
\end{align}

The drift term $f_i: \mathcal{A} \to \real_+$ in Equation (\ref{eq2}) is in $L^2$ space such that $\int_0^T f_i^2 ds < \infty$. $f_i$ is differentiable almost everywhere with respect to $a_i(t)$. The diffusion term can be interpreted as the principal's observations on each agent's actual actions contain noises and these noises may be intercorrelated. Let $\mathcal{F}^B_t$ represent the filtration generated by the Brownian motions $\pmb{B}(t)$ and $\pmb{B}_Z(t)$, and we assume that
the $m$ dimensional Brownian motions $\pmb{B}$ and $\pmb{B}_Z$ are independent. This is the extension of the output process proposed by Holmstrom and Milgram \cite{holmstrom1987aggregation} with most general direct interactions between $n$ agents.

Due to non-tractable actions, the principal offers agent $i$ a contract based on $X_i$ to indirectly influence his actions over the horizon. The agent then chooses strategy $\{a_i(t)\}_{0\leq t\leq T}$ that provides himself optimal returns. The principal also requires that these $n$ agents implement the desired actions concurrently. This is achieved by assuring that the agents' actions constitute a Nash equilibrium, which then disincentivizes the agent $i$ from deviating from the target action $a_i(t)$ for all $i$. An implementable and optimal dynamic contract must maximize the principal's utility $U_P$ under incentive-compatible (IC) and individual-rational (IR) constraints. We consider the case in which the participation constraint is only imposed at time zero. The IC-constraints are derived from analyzing agents' problems. The utility functions are as follows where $u_i: \mathcal{A}\times \mathcal{C}_i \to \real$ and $u_P:  \mathcal{A}\times \mathcal{C} \to \real$ are the instantaneous utility at time $t$ for agent $i$ and principal respectively\footnote{In what follows, $\pmb{1}$ is a $n$-dimensional identity vector. $\pmb{\Phi}(\pmb{Z}(T)) = [\Phi_1(Z_1(T)), \dots, \Phi_n(Z_n(T))]^{\intercal}$.}:  
\begin{align}
	U_i &=  \expt^{\pmb{a}} \left[ \int_0^T r e^{-rs} u_i( \pmb{a}(s),c_i(s)) d s + r e^{-rT}\Phi_i(Z_i(T)) \right] \\
    U_P &= \expt ^{\pmb{a}}\left[ \int_0^T r e^{-rs} \left[u_P(\pmb{a}(s),\pmb{c}(s)) \right] d s - r \pmb{1}^{\intercal} \cdot e^{-rT} \pmb{\Phi}(\pmb{Z}(T))  \right]
    \label{eq3}
\end{align}


In this work, all information about the processes and functions is private, and known only to the specific agents or principal, and the only shared information between agent $i$ and the principal is the agent $i$'s output $X_i(t)$ at time $t$. To assure that the actions of the agents constitute a Nash equilibrium, we assume that there is a third, neutral party who acts as a `messenger'.  The messenger has access to the agents actions, the principal's offered compensations and the processes and functions of each agent. It also has power to convince the agents that the actions indeed form a Nash equilibrium and thus must be adapted by the agents, The messenger informs the  principal of the actions of the agents at the equilibrium so that the principal can maximize $U_P$ (or the messenger can include this iteration within its algorithms). This is an implicit assumption in most multi-agent models and negotiation theory.  


\section{Individual Agent's Problem}\label{S3:SS1agent}
\noindent  In this section, we solve a single agent's problem given a fixed contract. Since the agents interact with the principal via individual contracts, we can adapt the framework and the methodology of Sannikov\cite{sannikov2008continuous} for the proofs of the results we obtain in this work \footnote{However, solving the individual agent problem does not mean that these contracts are mutually independent since the output process include the interactions among $n$ agents.}.

We assume that the instantaneous payoff the agent $i$ receives\footnote{Agent $i'$s return at $t$ from compensation $c_i(t)$ and the cost to agents for taking action $\pmb{a}(t)$.} is given by Equation (\ref{eq1}). 
Then the continuation payoff $W_i(t)$ is the agent $i$'s expected payoff received from $t$ to $T$ is defined as: 
\begin{align}
	W_i(t) &= \expt^{\pmb{a}} \left[ \int_t^T r e^{-r(s-t)} u_i( \pmb{a}(s),c_i(s)) d s + r e^{-r(T-t)}\Phi_i(Z_i(T))|\mathcal{F}_t^B \right] 
    \label{eq5}
\end{align}
  The two propositions that follow are proved in Sannikov in the infinite horizon case. The proofs carry over with almost no modifications for our case, even though we consider the finite horizon case with discounting. Thus we skip the proofs and refer the reader to  \cite{sannikov2008continuous}. 
\begin{proposition}\label{proposition1}
	For a fixed contract with finite utility for agent $i$, the dynamic evolution of his continuation payoff is determined by the SDE:
    \begin{align}
    	d W_i(t) = r\left( W_i(t) - u_i( \pmb{a}(t),c_i(t)) \right) d t + r Y_i(t)\sigma_idB(t)
     \label{eq6}
    \end{align}
    for some process $Y_i(t)$ that is adapted to $\mathcal{F}_t^B$. Conversely, $W_i(t)$ satisfying the SDE is the agent $i$'s continuation payoff.
\end{proposition}

Proposition \ref{proposition1} introduces a new process $Y_i(t)$ as ``the sensitivity of the agent's continuation payoff $W_i(t)$ to output $X_i(t)$'' by the martingale representation theorem. When solving the individual agent's problem, we can use the one-shot  deviation principle so that it is equivalent to the optimality of actions $\{\pmb{a}(t)\}_{0\leq t \leq T}$ to conditions on $\{Y_i(t)\}_{0 \leq t \leq T}$ as a series of static subproblems. Therefore, the principal only needs to send $\mathbf{Y}(t) = [Y_1(t), \dots Y_n(t)]^{\intercal}$ to the messenger when searching for equilibrium.

The key result in Proposition \ref{proposition1} is that an optimal contract can be written in terms of the state descriptor vector $(\pmb{W}(t),\pmb{Z}(t))$. The principal needs to specify these in individual contracts: (a) functions for the consumption process $c_i(\pmb{W}(t),\pmb{Z}(t))$ for each agent $i$; (b) functions of the target effort process $a_i(\pmb{W}(t),\pmb{Z}(t))$ for each agent $i$; (c) the laws of motion of continuation payoff $W_i(t)$ for all agents $i$ driven by output $\pmb{X}(t)$. An incentive compatible contract that provides consistent information for the three items and maximize the principals utility $U_P$ is optimal and implementable. The arguments for retirement at all $t\in [0, T]$ are still valid for multi-agent setting since one can take actions $0$ at $t$ (retire and $W_i(s) = 0$ for $s\geq t$) to guarantee himself a strictly positive $U_i$. For simplicity of notation, we denote a vector as $\pmb{x} = [x_1, \dots, x_i, \dots, x_n] = [\pmb{x}_{-i},x_i]$ in what follows.

\begin{proposition} \label{proposition2}
	For any fixed $\pmb{a}_{-i}(t)$, the contracted payment $c_i(t)$ for the agent $i$ is incentive-compatible if and only if strategy $\{ a_i(t) \}$ satisfies 
    \begin{align}
    	a_i(t) = \arg \max_{\tilde{a}_i(t) \in \mathcal{A}_i} \left[ Y_i(t) f_i(\pmb{a}_{-i}(t),\tilde{a}_i(t)) + u_i( \pmb{a}_{-i}(t),\tilde{a}_i(t),c_i(t)) \right]
    \label{eq7}
    \end{align}
    for all $t \in [0, T]$.
\end{proposition}

Without considering the Nash Equilibrium, agent $i$'s problem at time $t$ is to $\max_{a_i(t)} \expt \left[ U_i | \mathcal{F}^B_t\right]$. However, it is not sufficient for the principal to control process $W_i(t)$ to find the incentive-compatible contract $\{c_i(t), Y_i(t) \}$ separately as in the single-agent setting. Noticing that $u_i$ and $f_i$ are both functions of all agents decision variables, there are strong interactions among agents. To assure that $U_P$ is maximized, the principal needs to ensure that there is no agent $i$ who can benefit by deviating from the target $a_i(t)$. This can be done by choosing the target effort level $\pmb{a}(t)$to be a Nash Equilibrium at each time $t\in[0, T]$. In the next section, we present a set of sufficient conditions  for the existence of such an equilibrium.

\section{Subgame Perfect Nash Equilibrium}\label{S4:equilibrium}
\noindent In this section we prove the existence of a Nash Equilibrium for each given $\pmb{y}$ and $\pmb{c}$, which is incentive compatible, as defined by Equation (\ref{eq7}). This will be shown under the following assumptions on the functions $u_i$ and $f_i$ for all $i$:

\begin{enumerate}
	\item $u_i: \mathcal{A}\times \mathcal{C}_i\to \real$ is twice continuously differentiable, increasing in $\pmb{c}$, and concave.
    \item $f_i: \mathcal{A} \to \real_+$ is twice continuously differentiable, increasing and concave.
    \item For each $i$ and $\pmb{a}$, $f_i(\pmb{a}_{-i},a_i) \to \infty$ as $a_i \to \infty$, and $\frac{\partial f(\pmb{a})}{\partial a_i}\not= 0$ .
    \item The set $\cap_i\{(\pmb{a},\pmb{c}):u_i(\pmb{a},c_i)\geq 0 \mbox{ for all } i\} $ is nonempty and compact. 
    \item There exists an $m>0$ such that $m<\sup_xu_i(\pmb{a}_{-i},x,c_i)$, and $u_i\to -\infty$ as $x\to\infty$, for all $i$ and $\pmb{a}_{-i},c_i$.
    \item $u_i(\pmb{a}_{-i},0,c_i)\geq 0$ for each $\pmb{a}_{-i},c_i$.
\end{enumerate}

It is easy to check that Sannikov's single-agent contract is a special case where $u$ is separable with regard to $a(t)$ and $c(t)$ and $f(a(t)) = a(t)$. Assumption (4) is satisfied because an agent can take action $a_i(t) = 0$ and his utility will be $0$, which means the intersection of sets is not empty. Assumption (6) is valid by the same argument that $a_i(t) \notin \mathcal{A}_i$ if $u_i < 0$.  As a consequence of our Assumptions (1) - (6) above we can prove:
\begin{lemma} \label{lemma1}
Let $g_i^{\pmb{a}(t),c_i(t)}(x)=\dfrac{-u'(\pmb{a}_{-i}(t),x,c_i(t))}{f'(\pmb{a}_{-i}(t),x)}$. $g_i^{\pmb{a}(t),c_i(t)}$ is continuously differentiable and  monotone increasing as a function of $x$ in the domain $\mathcal{A}_i$. Also, there exist $0\leq\beta_i<\gamma_i$ such that for each $\beta_i<y<\gamma_i$ and $(\pmb{a},\pmb{c})\in \real^{2n}$, $g_i^{\pmb{a},c_i}(x)=y$ has a solution.
\end{lemma}

\proof{\bf{Proof of Lemma \ref{lemma1}:}}
Follows from properties (1) and (2) respectively of $u_i$ and $f_i$, is well defined from property (3) of $f'$, i.e., it is nonzero, and the property (1) of $u_i$ i.e., it is concave. Define $\gamma_i=\lim_{x\to\infty}\mbox{inf}_{\pmb{\alpha}\in \real^{n+1}}g_i^{\pmb{\alpha}}(x)$ which may be infinite, and $\beta_i=\mbox{max}\{0,\mbox{max}_{\pmb{\alpha}\in \real^{n+1}}g_i^{\pmb{\alpha}}(0)\}$. Let $y\in[\beta_i,\gamma_i]$ and define $\mbox{inf}_{\pmb{\alpha}}g_i^{\pmb{\alpha}}(\hat{x})=y$. Such an $\hat{x}$ exists since the function $\mbox{inf}_{\pmb{\alpha}}g_i^{\pmb{\alpha}}$ is monotone increasing. Now, $g_i^{\pmb{\alpha}}(\hat{x})\geq y$ and $g_i^{\pmb{\alpha}}(0)\leq \beta_i$ the result follows from the continuity of $g_i^{\pmb{\alpha}}$ and the intermediate value theorem.  $\rule{0.4em}{0.4em}$
\endproof

As a consequence of Lemma 1, define a set $ \mathcal{Y}=\prod_i [\beta_i,\gamma_i]$.

\begin{definition} \label{def1}
The agents effort $\pmb{a}$ is called a Nash Equilibrium if and only if deviation by one agent from the stipulated action in $\pmb{a}$ while the other agents follow their stipulated actions will result in a loss to the agent. This can be stated mathematically as: for each $i=1,\cdots,n$
\begin{align}
    	a_i \in \Gamma_i(\pmb{a}_{-i},c_i,y_i)=\{\hat{x}:\hat{x}= \arg \max_x \left[ y_i f_i(\pmb{a}_{-i},x) + u_i( \pmb{a}_{-i},x,c_i) \right]\}
    \label{eq8}
    \end{align}
  \end{definition}
  
We now prove a simple lemma to characterize the equilibrium:
\begin{lemma} \label{lemma2}
For almost all $t \in [0, T]$, and each $\pmb{y}(t)\in\mathcal{Y}$ and $\pmb{c}(t)\in\mathcal{C}$, if the Nash Equilibrium $\pmb{a}(t)$ exists, it lies in the set $\bigcap_i\{(\pmb{a}, c_i):u_i(\pmb{a},c_i)\geq 0\}$.
\end{lemma}

\proof{\bf{Proof of Lemma \ref{lemma2}:}}
For any given $\pmb{y}(t), \pmb{c}(t)$, let $\pmb{a}(t)$ be a Nash equilibrium for some $t\in[0,T]$ and let $u_i(\pmb{a}(s),c_i(s))<0$ for some $i$, and $s,t \in (t_1,t_2)$. Thus $\int_{t_1}^{t_2}u_i(\pmb{a}_{-i}(s),a_i(s),c_i(s))ds<0$. But, from property (6), $\int_{t_1}^{t_2} u_i((\pmb{a}_{-i}(s),0,c_i(s))ds\geq 0$. Thus $\pmb{a}(t)$ is not a Nash equilibrium, thus a contradiction. The Lemma \ref{lemma2} follows from the fact that $u_i(\pmb{a}_{-i}(t),a_i(t),c_i(t))<0$ on a set of measure 0 in $[0,T]$. $\rule{0.4em}{0.4em}$
\endproof
\vspace*{0.2em}

We show now that no agent has an incentive to leave before the terminal time if a Nash equilibrium for the actions of agents is adapted.
\begin{corollary}\label{corollary1:}
A consequence of the implementation of the Nash equilibrium is that no agent has an incentive to leave before the terminal time $T$.
\end{corollary}
\proof{\bf{Proof of Corollary \ref{corollary1:}}}
As is seen in the proof of Lemma \ref{lemma2} the Nash equilibrium gives the agent positive immediate return, thus making the total return an increasing function of the continuation time. $\rule{0.4em}{0.4em}$
\vspace*{0.2em}

We now establish the existence of a Nash equilibrium in the theorem as follows.
 
 \begin{theorem} \label{theorem1}
 For each given $\pmb{y}(t)\in\mathcal{Y}$ and $\pmb{c}(t)\in\mathcal{C}$, there exists a Nash Equilibrium $\pmb{a}(t)\in \mathcal{A}$. 
 \end{theorem}
 
 \proof{\bf{Proof of Theorem \ref{theorem1}:}}
 For a fixed agent $i$, given the concavity of the functions in Equation (\ref{eq6}), a necessary and sufficient condition for $\hat{x}$ to solve the optimization problem is that $g_i^{\pmb{a}(t),c_i(t)}(\hat{x})=y_i(t)$.  We note that as defined in Equation (\ref{eq8}) $\Gamma_i(\pmb{a}(t),c_i(t),y_i(t))=\{x:g_i^{\pmb{a}(t),c_i(t)}(x)=y_i(t)\}$. Now define a point-to-set map $\Gamma(\pmb{a}(t))=\Gamma^{\pmb{c}(t),\pmb{y}(t)}(\pmb{a}(t))=[\Gamma_1(\pmb{a}(t),c_1(t),y_1(t))),\cdots,\Gamma_n(\pmb{a}(t),c_n(t),y_n(t)))]$. Note that $\Gamma:\mathcal{A}\to\mathcal{A}^*$, where $\mathcal{A}^*$ is the set of all compact and convex subsets of $\mathcal{A}$. To see that $\Gamma$ is a upper hemi continuous point to set map, let $\pmb{a}^k$ be a sequence in $\mathcal{A}$ that converges to $\pmb{a}$. Also let $x^k\in\Gamma(\pmb{a}^k)$ for each $k$ such that $\pmb{x}^k$ converges to $\pmb{x}$. To see that $\pmb{x}$ is in $\Gamma(\pmb{a})$ we note that $x_i^k$ is such that $g_i^{\pmb{a}^k(t),c_i(t)}(x_i^k)=y_i(t)$. From the definition of $g_i$ in Lemma \ref{lemma1}, it is a continuous function of $\pmb{a}$, thus $y_i(t)=lim_{k\to\infty} g_i^{\pmb{a^k}(t),c_i(t)}(x^k)=g_i^{\pmb{a}(t),c_i(t)}(x)$ for each $i$. The existence of a Nash Equilibrium now follows from Lemma \ref{lemma2}, property (4) and the Kakutani Fixed Point Theorem.
 $\rule{0.4em}{0.4em}$
 \endproof

\section{The Optimum Contract}\label{S3:SS3principal}
\noindent In this section, we solve the principal's problem given that agents take actions at the Nash Equilibrium as in Section \ref{S4:equilibrium}. 

The principal's problem is as follows:
\begin{align}
U_P=\max_{\{ \pmb{c}(t),  \pmb{a}(t) \in \Theta(\pmb{c}(t),\pmb{y}(t)), 0 \leq t \leq T\} } \expt \left[ \int_0^T r e^{-rs} \left( u_P(\pmb{a}(\pmb{y}(s))),\pmb{c}(s) \right)ds - r e^{-rT}\pmb{1}^{\intercal} \cdot \pmb{\Phi}(\pmb{Z}(T))  \right]
\end{align}
where $\Theta(\pmb{c}(t),\pmb{y}(t))$ is the set of Nash equilibria for given $ \{ \pmb{c}(t), \pmb{y}(t) \}$.

Let the solution $\pmb{c}^*,\pmb{y}^*$ to the principal's problem exist and define the principal's optimum return from some $t>0$ to $T$ as

\begin{align}
R_P(t)=\max_{\{\pmb{c}(s),\pmb{y}(s), s\geq t\}}\expt^{[f(\pmb{a})]}\left[ r\int_t^Te^{-r(u-t)} \left( u_P(\pmb{a}(\pmb{y}(u)),\pmb{c}(u) \right)du-r e^{-r(T-t)}\pmb{1}^{\intercal} \cdot \pmb{\Phi}(\pmb{Z}(T))|\mathcal{F}_t^B\right]
\end{align}
and note that the solution as defined, is $\pmb{c}^*, \pmb{y}^*$. We now make an

\begin{assumption}
We assume that the value $R_p(t)$ has the following $C^{1,2,2}$  functional\footnote{$F^P$ is once continuously differentiable in t and twice in the other two variables.} form $F^P(t,\pmb{W}(t),\pmb{Z}(t))$ of $t$, the agents' continuation vector $\pmb{W}(t)$ and the termination value descriptor vector $\pmb{Z}(t)$\footnote{In what follows, for the ease of exposition we will shorten $F^P(t,\pmb{W}(t),\pmb{Z}(t))$ to $F^P_t$ whenever there is no possibility of confusion.}. 
\end{assumption}

Thus the optimum value received by the principal is given by
\begin{align}
U_P(t)=r\int_0^t e^{-rs} \left( u_P(\pmb{a}(\pmb{y^*}(s)),\pmb{c}^*(s)) \right)ds+e^{-rt}F^P(t,\pmb{W}(t),\pmb{Z}(t))
\label{eq10}
\end{align}

It is easy to show (see Sannikov \cite{sannikov2008continuous}) that $U^P(t)$ defined by Equation (\ref{eq10}) is a Martingale, and thus has zero drift. Applying Ito's multidimensional lemma and the dynamics of $\pmb{W}(t)$ and $\pmb{Z}(t)$, we obtain the dynamics of $U_P(t)$, and then the HJB formulation of the principals problem by setting the drift of the dynamics to 0. Using the dynamics of $W_i(t)$ as in Equation (\ref{eq6}), we get
\begin{align}
d\pmb{W}(t)= r\left[ \pmb{W}(t)-\pmb{u}(\pmb{a}(\pmb{y}(t)),\pmb{c}(t)) \right]dt+ rY(t) \sigma dB(t)
\end{align}
and similarly using Equation (\ref{eqn11}), for $\pmb{Z}(t)$.
Here $\sigma$ and $\sigma_Z$ are  $n\times m$ matrices and $Y(t)=\mbox{diag}(\pmb{y}(t))$. Let the differential operator $\mathcal{H}^{\pmb{y}}$ with $\mu(t,\pmb{W}(t))=\pmb{W}(t)-\pmb{u}(\pmb{a}(\pmb{y}(t)),\pmb{c}(t))$ and $\sigma(\pmb{y})=Y\sigma$ be defined as\footnote{In the formula $D_xF$ is the Jacobean and $ D^2_xF$ is the Hessian of the $C^2$ function $F$ of $x$}:
\begin{eqnarray}
\mathcal{H}^{\pmb{y}}F^P_t&=&r D_{\pmb{w}}F^P_t\mu(t,\pmb{w}(t))+r D_{\pmb{z}}F^P_t\mu_{\pmb{z}}(\pmb{a}(t),\pmb{c}(t))+ \nonumber \\&&\frac{1}{2}r^2\mbox{trace}(\sigma(\pmb{y}(t))^{\intercal}D^2_{\pmb{w}}F^P_t\sigma(\pmb{y}(t)))+\sigma_{\pmb{z}}^TD^2_{\pmb{z}}F^P_t\sigma_{\pmb{z}})
\end{eqnarray}

Applying the multidimensional Ito's lemma, we get the drift of the dynamics of $U_P(t)$ as
\begin{align}
e^{-rt}\left[\frac{\partial}{\partial t}F^P_t+r u_P(\pmb{a}(\pmb{y}^*(t)),\pmb{c}^*(t))+\mathcal{H}^{\pmb{y}}F^P_t-rF^P_t\right]
\end{align}

Thus the Hamilton-Jacobi-Bellman (HJB) equation that results is:
\begin{eqnarray}
\frac{\partial}{\partial t}F^P_t+\max_{\pmb{y},\pmb{c}}\left\{r u_P(\pmb{a}(\pmb{y}(t)),\pmb{c}(t))+\mathcal{H}^{\pmb{y}}F^P_t \right\}-r F^P_t &=& 0\\
F^P(T,\pmb{w},\pmb{z})&=&-r \pmb{1}^{\intercal} \cdot\pmb{\Phi}(\pmb{z}) \mbox{ for all } \pmb{w}, \pmb{z} \\
\pmb{a}(\pmb{y}(t),\pmb{c}(t)) &\in& \Theta(\pmb{c}(t),\pmb{y}(t))\mbox{ for all } t 
\label{HJB1}
\end{eqnarray}

One can now use standard arguments to show that the stipulated function $F^P(t,\pmb{w},\pmb{z})$ does indeed generate the optimal policy for the multi-agent problem. We omit this verification theorem here.

\section{Solving the principal-multi-agent problem} \label{S3:solve}
\begin{figure}[!ht]
	\centering
    \includegraphics[width = \textwidth]{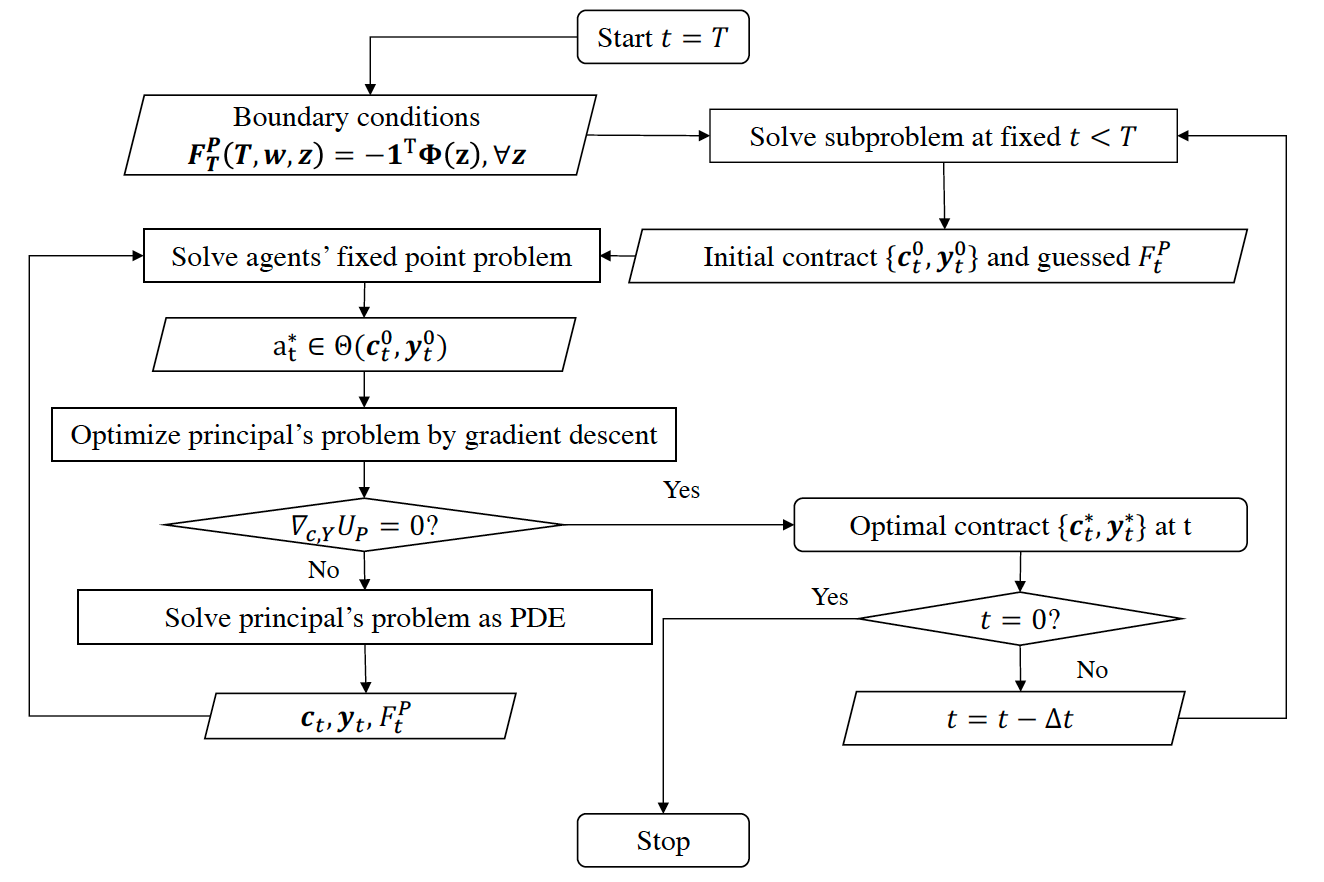}
    \caption{Flowchart to solve the optimal contracts.}
    \label{figure1}
\end{figure}

\noindent The framework to solve the single-principal multi-agent problem is shown in Figure \ref{figure1}. 
Solving the contracts adopt the backward scheme in time horizon. Fixing an arbitrary point $(t, \pmb{w},\pmb{z})$, and an arbitrary guess of the function $F^P_t$,  we solve a fixed point problem,  \cite{eavessaigal72}, to find the Nash Equilibrium of $n$-agents' actions for given contracts $\{\pmb{c}^0, \pmb{y}^0 \}$. Then we solve a static optimization to find the optimal contracts iteratively by possibly a gradient descent method in principal's problem. The function is then updated by solving the PDE for the unknown function $F^P$. Unfortunately there is no general analytic method to accelerate this guess-and-check process. In most cases, we can only solve for some special functions, some examples of which include quadratic utility functions of state $\pmb{w}(t),\pmb{z}(t)$. One observation in the literature speculates that $F^P$ appears to inherent some structural properties from $u_P$ and $\pmb{\Phi}$\cite{bjork2009arbitrage}. In case that there is no explicit solution to PDE, we can use numerical methods to solve the parabolic PDE.

\section{Conclusion}\label{S5:conclusion}
\noindent In this paper, we prove the existence of optimal contracts in the multi-agent version of continuous-time dynamic moral hazard. We are able to show that there is sub-game perfect Nash equilibrium in a multi-agent system by imposing certain information, compactness and convexity conditions. Similar to the single-agent model, we can use dynamic programming to solve the optimal contracts as a series of static optimizations. The principal's problem is converted to HJB equation. Furthermore, a backward iteration algorithm to find the optimal contracts is provided. 

In a future work, we will consider special cases and find the optimal contracts (possibly via simulation) to obtain some understanding of the effect on the optimal contracts of the interactions between different types of agents. This could give some additional insights beyond ones obtained from a single agent model. We expect to simulate continuous time moral hazard situations in traditional areas like finance and labor markets, and possibly new areas like energy and transportation.


%
%
%

\section*{Acknowledgments.}
We thank Heng Liu, Robert Hampshire, Reza Kamaly, Abdullah Alshelahi, Jingxing Wang who provided insight and expertise that greatly assisted the research.


\bibliographystyle{informs2014} 
\bibliography{reference.bib} 


\end{document}